\documentstyle[epsf,12pt]{article}
\newcommand {\beq}{\begin{equation}}
\newcommand {\eeq}{\end{equation}}
\newcommand {\beqa}{\begin{eqnarray}}
\newcommand {\eeqa}{\end{eqnarray}}
\newcommand {\n}{\nonumber \\}

\renewcommand{\theequation}{\thesection.\arabic{equation}}
\begin{document}
\setlength{\oddsidemargin}{0cm}
\setlength{\baselineskip}{7mm}

\begin{titlepage}
 \renewcommand{\thefootnote}{\fnsymbol{footnote}}
$\mbox{ }$
\begin{flushright}
\begin{tabular}{l}
KEK-TH-1002\\
Dec. 2004
\end{tabular}
\end{flushright}

~~\\
~~\\
~~\\

\vspace*{0cm}
    \begin{Large}
       \vspace{2cm}
       \begin{center}
{Wilson Line Correlators in ${\cal N}=4$\\ 
Non-commutative Gauge Theory on $S^2 \times S^2$}
\\
       \end{center}
    \end{Large}

  \vspace{1cm}

\begin{center}
           Yoshihisa K{\sc itazawa}$^{1),2)}$\footnote
           {
e-mail address : kitazawa@post.kek.jp}
           Yastoshi T{\sc akayama}$^{2)}$\footnote
           {
e-mail address : takaya@post.kek.jp}\\{\sc and}
           Dan T{\sc omino}$^{1)}$\footnote
           {
e-mail address : dan@post.kek.jp}

        $^{1)}$ {\it High Energy Accelerator Research Organization (KEK),}\\
               {\it Tsukuba, Ibaraki 305-0801, Japan} \\
        $^{2)}$ {\it Department of Particle and Nuclear Physics,}\\
                {\it The Graduate University for Advanced Studies,}\\
{\it Tsukuba, Ibaraki 305-0801, Japan}\\
\end{center}

\vfill

\begin{abstract}
\noindent
We investigate the Wilson line correlators  
dual to supergravity multiplets in ${\cal N}=4$
non-commutative gauge theory on $S^2\times S^2$.
We find additional non-analytic contributions to the correlators
due to UV/IR mixing in comparison to ordinary gauge theory.
Although they are no longer BPS off shell,
their renormalization effects are finite as long as they carry
finite momenta.
We propose a renormalization procedure to obtain local operators
with no anomalous dimensions in perturbation theory. 
We reflect on our results from dual supergravity point of view.
We show that supergravity can account for both IR and UV/IR contributions.
\end{abstract}
\vfill
\end{titlepage}
\vfil\eject

\section{Introduction}
\setcounter{equation}{0}
\setcounter{footnote}{0}

It is a very attractive idea to obtain string theory in the large $N$ limit
of gauge theory \cite{tHooft}. 
The most concrete proposal so far is $AdS$/CFT correspondence 
\cite{Maldacena}. The correspondence between the BPS sectors of the both theories 
is well established \cite{GKP}\cite{Witten}.
The correspondence is extended further to (close to BPS) BMN operators \cite {BMN}
and to integrable sectors \cite{MZ}$\sim$\cite{AS}.
Since the correspondence is the duality between the weak and strong coupling regimes,
we presumably require non-perturbative formulation of supersymmetric gauge theories to
make further progress.

In fact matrix models have been proposed to formulate string theory at the non-perturbative level
\cite{BFSS}\cite {IKKT}.
The most attractive feature of these constructions is that they preserve supersymmetry.
Non-commutative (NC) gauge theory can be realized by expanding the matrices around
the flat non-commutative solutions in the large $N$ limit \cite{CDS}\cite{AIIKKT}\cite{Li}. 
A fuzzy homogeneous space $G/H$ can be obtained as a classical solution by introducing a Myers term
\cite{Myers}\cite{IKTW}\cite{Mathom} with finite $N$.
It may extremize the effective action of IIB matrix model at quantum level \cite{fuzS2S2}\cite{IT}.
Although SUSY is broken softly in these models at the scale where the manifold
is curved, it will not affect local properties of the theory .
In this sense non-perturbative formulation of supersymmetric NC gauge theory
may be realized through matrix models.

Just like ordinary gauge theory, supersymmetric NC gauge theory is expected 
to possess dual supergravity description \cite{hashimoto}\cite{ads+f}.
In fact there are BPS operators in matrix models which serve as the vertex operators
for supergravity multiplets \cite{vertex}\cite{ITU}. In a non-commutative spacetime, 
they reduce to a special type of the Wilson lines which are the only gauge invariant operators
in NC gauge theory \cite{IIKK}. 
However they are no longer BPS operators off-shell unlike ordinary gauge theory.
In fact they are renormalized in general and the justification of supergravity description requires
more work than ordinary gauge theory. 
They are referred to as SUGRA operators in this paper.
The main purpose of this paper is to advance our understandings in this issue.

We investigate non-commutative extensions of chiral operators in ${\cal N}=4$ NC gauge theory 
which is realized on fuzzy $S^2\times S^2$ in the large $N$ limit.
Since compact spaces can be realized in matrix models with finite $N$,
such a construction may enable us to understand non-perturbative (finite $N$)
effects in string theory.
We first construct these operators in the matrix model
in section 2.
We subsequently investigate the correlators of them
in perturbation theory in section 2 and 3.
The two point correlators receives two types of contributions.
The first type is positive definite just like ordinary gauge theory
while the second type contains non-commutative phases.
We call them planar and non-planar contributions respectively in this paper.
In the literature, it has been often assumed that the correlators reduce
to those of ordinary gauge theory in the small momentum limit.
However that needs to be examined due to UV/IR mixing effects \cite{MRS}.
In fact we have found that the correlators receive UV/IR contributions
in addition to IR contributions which are the sole contributions in ordinary gauge theory
\footnote{
The large momentum limit of the Wilson line correlators
has been investigated in \cite{Gross}\cite{DhKhe}\cite{RR}.}.

Non-planar contributions also play an important role to understand
the renormalization property of the SUGRA operators.
Since NC gauge theory is not a local field theory, the introduction of the
notion of locality requires a considerable work at quantum level. 
We find in section 3 that 
the Wilson line correlators receive logarithmic corrections at the one loop level.
We recall here that the Wilson lines with different momenta are different operators
and there is a freedom to rescale them by momentum dependent factors. 
Since the renormalization effect can be associated with each
Wilson line, we propose a perturbative prescription to rescale
the operators in such a way that
they can be interpreted as the Fourier 
transformation of the local operators with no anomalous dimensions.

Such a prescription is certainly necessary to make contact with supergravity.
We investigate dual supergravity description of the
correlators in section 4. 
The precise prescription to apply such a correspondence is not fully
understood such as where to locate the Wilson lines in the fifth radial coordinate. 
In this paper we study SUGRA operators in detail in order to understand
this problem.
We show that we can successfully reproduce the essential features of the
correlators in NC gauge theory by locating the Wilson lines at the maximum
of string metric as proposed in \cite{DhK}.
We conclude in section 5 with discussions.

\section{Wilson lines on homogeneous spaces}
\setcounter{equation}{0}

NC gauge theory on homogeneous space $G/H$ 
has been investigated in our recent work \cite{fuzS2}$\sim$\cite{KTT}.
In this construction, a fuzzy spacetime is represented by $p_{\mu}$ which 
is a Lie algebra of $G$.
$p_{\mu}$ and gauge field $a_{\mu}$
are unified into the basic matrix degrees of freedom as $A_{\mu}=f(p_{\mu}+a_{\mu})$.
$f$ is the coefficient of a Myers term.
The 't Hooft coupling is identified as $\lambda^2\sim n^2/f^4N$
in 2 and 4 dimensional NC gauge theories with $U(n)$ gauge group. 
We have clarified the large $N$ scaling behavior of the theory by power counting
arguments with fixed 't Hooft couplings \cite{fuzS2S2}.
These predictions for the large $N$ scaling behavior of matrix models are confirmed by 
recent non-perturbative investigations \cite{ABNN1}\cite{ABNN2}\cite{ANN}.

In this section, we first construct the gauge invariant observables, namely
the Wilson lines in NC gauge theory on $G/H$.
They are the single traced object made of polynomials of matrices. 
The structure of these observables is dictated by the isometry $G$.
On $S^2$, we consider the following polynomials of $A_{\mu}$
\beq
y_{jm}^{\alpha_1,\alpha_2,\cdots,\alpha_j}TrA_{\alpha_1}A_{\alpha_2}\cdots A_{\alpha_j}
A_{i_1}A_{i_2}\cdots A_{i_k} ,
\eeq
where $\alpha =0,1,2$ denote the dimensions in which $S^2$ extends.
$y_{jm}^{\alpha_1,\alpha_2,\cdots,\alpha_j}$ denotes
a totally symmetric traceless tensor which corresponds to the spin $j $ representation of $SU(2)$.
The one point functions of these observables vanish since they carry non-vanishing
angular momentum. 
On $S^2\times S^2$, we can construct analogous operators
\beq
y_{jm}^{\alpha_1,\alpha_2,\cdots,\alpha_j}
y_{pq}^{\beta_1,\beta_2,\cdots,\beta_p}
TrA_{\alpha_1}A_{\alpha_2}\cdots A_{\alpha_j}
A_{\beta_1}A_{\beta_2}\cdots A_{\beta_p}
A_{i_1}A_{i_2}\cdots A_{i_k} , 
\eeq
where $\beta=3,4,5$ denote the dimensions in which the second
$S^2$ extends.

Since $i=6\sim 9$, these operators are classified by
the representations of $SO(4)$ or its subgroup.
The simplest operators of this kind possess the $U(1)$(R) charge 
which is equal to
the number of the $Z$ fields in the operator:
\beq
y_{jm}^{\alpha_1,\alpha_2,\cdots,\alpha_j}
y_{pq}^{\beta_1,\beta_2,\cdots,\beta_p}
TrA_{\alpha_1}A_{\alpha_2}\cdots A_{\alpha_j}
A_{\beta_1}A_{\beta_2}\cdots A_{\beta_p}
Z^J  ,
\eeq
where $Z=(A_8+iA_9)/\sqrt{2}$.
Although this operator is the analogue of the chiral operator in ordinary
gauge theory, it is not invariant under SUSY transformation of IIB matrix model.
It is due to the presence of gauge fields in addition to $Z^J $
which carry definite momenta. 
In contrast we do not need such structure in ordinary gauge theory. 
Therefore this operator will be renormalized in generic cases.
The main goal of this paper is to understand the renormalization
property of SUGRA operators in the small momentum regime.

In the case of $S^2$, the  background $p_{\mu}$ consists of angular momentum operators
in the spin $l$ representation. In this paper we focus on a simple 4d manifold
$S^2\times S^2$ where the both $S^2$ are of the identical 
size: $l_1=l_2=l$ and $N=n(2l+1)^2$ with $U(n)$ gauge group.
In our expansion of $A_{\mu}$ around a background $p_{\mu}$,  the leading term of the Wilson line is:
\beq
f^{j+p+J}
y_{jm}^{\alpha_1,\alpha_2,\cdots,\alpha_j}
y_{pq}^{\beta_1,\beta_2,\cdots,\beta_p}
Trp_{\alpha_1}p_{\alpha_2}\cdots p_{\alpha_j}
p_{\beta_1}p_{\beta_2}\cdots p_{\beta_p}z^J ,
\eeq
where $z=(a_8+ia_9)/\sqrt{2}$. 
Without the loss of generality, we can focus on the highest weight states of $SU(2)\times SU(2)$:
\beq
y_{j,j}y_{p,p}Tr(p_{+})^{j}(\tilde{p}_{+})^{p}z^J
=Tr{\cal Y}_{j,p}z^J ,
\label{yz}
\eeq
where we have also rescaled the operator to absorb the factor of $f^{j+p+J}$.

Since we normalize $Tr{\cal Y}_{j,p}^{\dagger}{\cal Y}_{j,p}=n$,
the coefficient $y_{j,j}$ is determined semiclassically:
\beq
tr(p_{+})^{j}(p_{-})^{j}
\sim ({1\over 2})^{j}tr(p_1^2+p_2^2)^j
\sim ({l^2\over 2})^{j}{l}
\int dcos\theta  sin^{2j}\theta 
\eeq
as
\beq
y_{j,j}^2={(2j+1)!!\over (2j)!!}\Big({2\over l^2}\Big)^{j}{1\over 2l} .
\eeq
If we replace one of $p_{\mu}$ by a gauge field $a_{\mu}$,
we essentially obtain an operator with a different spherical harmonics ${\cal Y}_{j-1,p}$.
Since $y_{j,j}\sim l^{-(j+1/2)}$, such an operator
is suppressed by $1/l$ in comparison to the original one.
Thus the non-leading terms which contain gauge fields instead of $p_{\mu}$ are
suppressed by powers of $1/l$ in comparison to the leading term.
We therefore neglect them in the subsequent investigations.
We will indeed find in section 3
that they can be neglected in the large $N$ limit with a fixed 't Hooft coupling $\lambda^2$
as long as $j,p$ are $O(1)$.
However their effect becomes important when $j^2+p^2\sim l/\lambda$.  
It is expected the case since the momenta approaches the non-commutative scale. 
In this paper, we assume that momenta $j,p$ of the Wilson lines are
much smaller than the non-commutative scale .

We first investigate the two point functions of $Tr{\cal Y}_{j,p}z^2$ with $J=2$
which correspond to the most relevant chiral operator in ordinary $SU(n)$ gauge theory. 
At the tree level, we obtain both the planar and non-planar contributions as follows
\beqa
&&{n\over N}<Tr{\cal Y}_{j,p}z^2Tr\bar{z}^2{\cal Y}_{j,p}^{\dagger}>\n
&=&\epsfysize=15mm\parbox[c]{25mm}{\epsfbox{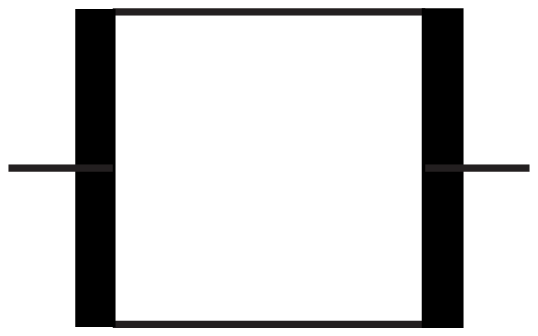}}
+\epsfysize=15mm\parbox[c]{25mm}{\epsfbox{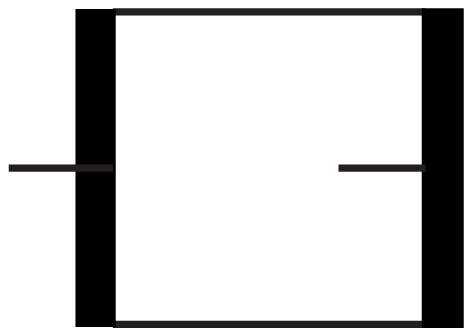}}\n
&=&<j,p|{1\over P_2^2P_3^2}|j,p>_p
+<j,p|{1\over P_2^2P_3^2}|j,p>_{np},
\label{2jcor}
\eeqa
where 
\beqa
P_i^{\mu}{\cal Y}_{j_{i'}m_{i'}p_{i'}q_{i'}}&\equiv&  [p^{\mu},{\cal Y}_{j_{i'}m_{i'}p_{i'}q_{i'}}]
\delta_{ii'},
\n
{\cal Y}_{jmpq}&\equiv& y_{jm}^{\alpha_1,\alpha_2,\cdots,\alpha_j}
p_{\alpha_1}p_{\alpha_2}\cdots p_{\alpha_j}
y_{pq}^{\beta_1,\beta_2,\cdots,\beta_p}
p_{\beta_1}p_{\beta_2}\cdots p_{\beta_p}.
\eeqa
We have also introduced the following average:
\beqa
<j,p|X|j,p>_p
&=&{n^3\over f^8N}\sum_{j_2,j_3,m_2,m_3}\sum_{p_2,p_3,q_2,q_3}
\Psi_{123}^*X \Psi_{123},\n
<j,p|X|j,p>_{np}
&=&{n^3\over f^8N}\sum_{j_2,j_3,m_2,m_3}\sum_{p_2,p_3,q_2,q_3}
\Psi_{132}^*X \Psi_{123},\n
\Psi_{123}&\equiv &
Tr {\cal Y}_{j_3m_3p_3q_3}{\cal Y}_{j_2m_2p_2q_2}{\cal Y}_{jp}.
\label{avr1}
\eeqa

The planar amplitude is
\beqa
<j,p|{1\over P_2^2P_3^2}|j,p>_p&=&{n^3\over f^8N}\sum_{j_2,j_3,p_2,p_3}
{(2j_2+1)(2p_2+1)(2j_3+1)(2p_3+1)\over
(j_2(j_2+1)+p_2(p_2+1))(j_3(j_3+1)+p_3(p_3+1))}\n
&&\times     \left\{
 \begin{array}{ccc}
  j &  j_2 & j_3 \\
  l   &  l & l
 \end{array}
\right\}^2
     \left\{
 \begin{array}{ccc}
  p &  p_2 & p_3 \\
  l   &  l & l
 \end{array}
\right\}^2 ,
\label{plam2j}
\eeqa
where we refer \cite{Edm} for $6j$ symbols.
Apart from a numerical factor, this function is identical to $\omega (P^2)$ 
which appeared in the one loop self-energy of gauge fields \cite{KTT}.
The non-planar amplitude is
\beqa
<j,p|{1\over P_2^2P_3^2}|j,p>_{np}&=&{n^3\over f^8N}\sum_{j_2,j_3,p_2,p_3}
{(2j_2+1)(2p_2+1)(2j_3+1)(2p_3+1)\over
(j_2(j_2+1)+p_2(p_2+1))(j_3(j_3+1)+p_3(p_3+1))}\n
&&\times e^{i\phi_{123}}      \left\{
 \begin{array}{ccc}
  j &  j_2 & j_3 \\
  l   &  l & l
 \end{array}
\right\}^2
     \left\{
 \begin{array}{ccc}
  p &  p_2 & p_3 \\
  l   &  l & l
 \end{array}
\right\}^2 ,
\eeqa
where $e^{i\phi_{123}}= (-1)^{j+j_2+j_3+p+p_2+p_3}$.
Note that this amplitude is planar  with respect to the gauge group indices.
We investigate only planar sectors with respect to the gauge group indices
by assuming $n$ is large in this paper.
Unlike large momentum regime, it cannot be neglected in comparison to 
the planar amplitude (\ref{plam2j})
in the small momentum regime.  It exhibits extra non-analytic behavior
with respect to the external momenta due to UV/IR mixing
as we shortly demonstrate.

A detailed investigation of the planar amplitude using the Wigner
approximation of $6j$ symbols have been carried out in \cite{KTT}.
Such an approximation can be justified for large external momenta
$j^2+p^2>>1$.
Here we investigate these amplitudes
by using Edmonds' approximation for 6j symbols:
\beq
\left\{
\begin{array}{ccc}
  j &  j_2 & j_3 \\
  l   &  l & l
 \end{array}\right\}
 ={(-1)^{j_3}\over \sqrt{(2j_3+1)(2l+1)}}d_{j_3-j_2,0}^{(j)}(\theta_3 ) ,
\eeq
where
\beq
cos(\theta_3 )=-{1\over 2}\sqrt{j_3(j_3+1)\over l(l+1)} ,
\label{Edm}
\eeq
and
\beq
d_{m'm}^{(j)}(\beta )=(jm'|exp({i\beta J_y})|jm) .
\eeq
Since it is valid when $j_2,j_3 >>j$, this approximation enables us
to estimate logarithmically divergent amplitudes with finite external momenta .

Under this approximation, the planar amplitude becomes
\beqa
&&({n^2\over f^4N})^2\sum_{j_2,j_3,p_2,p_3}
{(2j_2+1)(2p_2+1)\over
(j_2(j_2+1)+p_2(p_2+1))(j_3(j_3+1)+p_3(p_3+1))}\n
&&\times (d_{j_3-j_2,0}^{(j)}(\theta_3 ))^2(d_{p_3-p_2,0}^{(p)}(\tilde\theta_3 ))^2 .
\eeqa
Since $|j_3-j_2|\leq j,|p_3-p_2|\leq p$, we may approximate the above
\beqa
&&({n^2\over f^4N})^2\sum_{j_2,j_3,p_2,p_3}
{(2j_3+1)(2p_3+1)\over
(j_3(j_3+1)+p_3(p_3+1))^2}
(d_{j_3-j_2,0}^{(j)}(\theta_3 ))^2(d_{p_3-p_2,0}^{(p)}(\tilde\theta_3 ))^2\n
&=&({n^2\over f^4N})^2\sum_{j_3,p_3}
{(2j_3+1)(2p_3+1)\over
(j_3(j_3+1)+p_3(p_3+1))^2}\n
&\sim&({n^2\over f^4N})^2log({4l^2\over j^2+p^2}) ,
\label{J2pl}
\eeqa
where the lower cut-off is provided by the external  momenta.

The non-planar amplitude can be estimated under the same approximation as
\beqa
&&({n^2\over f^4N})^2\sum_{j_2,j_3,p_2,p_3}e^{i\phi_{123}}
{(2j_2+1)(2p_2+1)\over
(j_2(j_2+1)+p_2(p_2+1))(j_3(j_3+1)+p_3(p_3+1))}\n
&&(d_{j_3-j_2,0}^{(j)}(\theta_3 ))^2(d_{p_3-p_2,0}^{(p)}(\tilde\theta_3))^2 .
\eeqa
In an analogous way, it can be evaluated as
\beqa
&&({n^2\over f^4N})^2\sum_{j_2,j_3,p_2,p_3}
{(2j_3+1)(2p_3+1)\over
(j_3(j_3+1)+p_3(p_3+1))^2}e^{i\phi_{123}}
(d_{j_3-j_2,0}^{(j)}(\theta_3 ))^2(d_{p_3-p_2,0}^{(p)}(\tilde\theta_3))^2\n
&=&({n^2\over f^4N})^2\sum_{j_3,p_3}
{(2j_3+1)(2p_3+1)\over
(j_3(j_3+1)+p_3(p_3+1))^2}
d_{0,0}^{(j)}(\pi-2\theta_3 )d_{0,0}^{(p)}(\pi-2\tilde\theta_3 )\n
&=&({n^2\over f^4N})^2\sum_{j_3,p_3}
{(2j_3+1)(2p_3+1)\over
(j_3(j_3+1)+p_3(p_3+1))^2}
P_j\Big(1-{j_3(j_3+1)\over 2l(l+1)}\Big)
P_p\Big(1-{p_3(p_3+1)\over 2l(l+1)}\Big) .\n
\eeqa
Since the Legendre polynomials $P_j(cos\theta )$ oscillate, we may identify the upper cut-off
of the summations with the location of their first node.
From the asymptotic behavior of the Legendre polynomials for large $j$:
\beq
P_j(cos\theta )\sim \sqrt{2\over j\pi sin\theta}sin\Big((j+{1\over 2})\theta+
{\pi\over 4}\Big) ,
\eeq
we can estimate the location of their first node as $\theta\sim \pi /j$.
We thus find the upper cut-off of the summations as $j_3<l/j$ and $p_3<l/p$.
In this way, we obtain
\beq
({n^2\over f^4N})^2\int_{j^2}^{({l\over j})^2} dx\int_{p^2}^{({l\over p})^2} dy
{1\over (x+y)^2}
\sim {\lambda^4\over (4\pi)^4} log ({l^2\over (j^2+p^2)^2}) ,
\label{J2np}
\eeq
where $\lambda^2=(4\pi)^2n^2/f^4N$ has been identified with the 
't Hooft coupling \cite{fuzS2S2}.

The corresponding Wilson line of the NC gauge theory in the flat 4d space would be
\beq
Trexp({1\over l}ik\cdot A)Z^2 .
\eeq
Let us compute the corresponding two point function with the identical UV cut-off $l$.
The planar amplitude would behave as
\beq
{\lambda^4\over (2\pi)^4}
\int^l d^4q {1\over q^2 (k+q)^2}
\sim {\lambda^4\over (4\pi)^2} log {4l^2\over k^2} ,
\label{pldiv}
\eeq
while that of the non-planar amplitude behaves as
\beqa
&&{\lambda^4\over (2\pi)^4}
\int^l d^4q {1\over q^2 (k+q)^2}exp({1\over l}iq\cdot k)\n
&=&{\lambda^4\over (4\pi)^2}\Big(\int_{k^2}^{l} dq^2 {1\over q^2}+
\int_{l}^{l^2\over k^2} dq^2 {1\over q^2}\Big)\n
&=&{\lambda^4\over (4\pi)^2}\Big(log({l\over k^2})+ log({l\over k^2})\Big) .
\label{npldiv}
\eeqa
We observe that the non-analytic behavior of the correlators 
(\ref{J2pl}) and (\ref{J2np}) on $S^2\times S^2$ and 
(\ref{pldiv}) and (\ref{npldiv}) on the flat 4d space are identical
with the identification of the external momenta as $j^2+p^2\sim k^2$. 
This coincidence is expected to hold in the large $N$ limit
where $S^2\times S^2$ becomes locally flat.
The correlators on $S^2\times S^2$ and
the flat 4d space should agree as long as their momenta are large enough to probe a local
region which is indistinguishable in the both cases.

In (\ref{npldiv}), we have separated the integral into small and large
momentum contributions.
The lower cut-off of the integral is provided by the external momenta.
The small momentum contribution
can be associated with the Wilson coefficient of
the leading OPE expansion
since the derivations with respect to the external momenta
render it finite.
On the other hand, the upper cut-off of the integral comes
from the rapidly oscillating phase of the integrand. 
Due to the uncertainty relation in non-commutative space
$\Delta x \Delta y \sim l$ and $\Delta x \Delta k \sim l$, a quantum which carries
momentum larger than $\sqrt{l}$ extend the same amount
in the perpendicular direction. Therefore the large momentum
contribution
can be interpreted in the dual coordinate space
where the upper cut-off ${l/ k}$ can be identified as a long distance cut-off. 
In this way we can identify it as the Fourier transformation of the
long rang interaction ${1/ x^4}$ as
\beq
{\lambda^4\over (2\pi)^4}\int_{\sqrt{l}} {d^4 x}{1\over x^4}exp({1\over l}ik\cdot x) .
\label{lrint}
\eeq

We thus conclude that the two point function of the Wilson lines with $J=2$
(\ref{2jcor}) behaves as follows for small external momenta to the leading order
in perturbation theory
\beq
{\lambda^4\over (4\pi)^4}\Big(2log(l/P^2)+log(l/P^2)+log(l)\Big) .
\label{2jtree}
\eeq
The first term is identical to ordinary gauge theory 
if we identify the UV cut-off with non-commutative scale $\sqrt{l}$.
The second term indicates a long range interaction due to UV/IR mixing
specific to NC gauge theory as in (\ref{lrint}).
We discard the last term as it corresponds to $\delta$ function in coordinate space.
This long range interaction is consistent with that of a Kaluza-Klein
mode which couples to this operator in supergravity interpretation.
The first term can be interpreted in terms of the Kaluza-Klein mode just like
ordinary AdS/CFT correspondence. Nevertheless
the second term represents the extra long range interaction which is absent
in ordinary gauge theory.
We investigate the supergravity description of NC gauge theory correlators
in section 5.

The two point functions of generic chiral operators with $J> 2$ behave at tree level as
\beqa
&&{n\over N}<Tr{\cal Y}_{j,p}z^JTr\bar{z}^J{\cal Y}_{j,p}^{\dagger}>\n
&=&<j,p|\prod_{i=1}^J{1\over P_i^2}|j,p>_p
+<j,p|\prod_{i=1}^J{1\over P_i^2}|j,p>_{np} ,
\label{Jtree}
\eeqa
where the planar amplitude is
\beqa
<j,p|\prod_{i=1}^J{1\over P_i^2}|j,p>_p
&\equiv&{n^{J+1}\over f^{4J}N}
\sum_{2\cdots J}\Phi_{12\cdots J}^*
\Big(\prod_{i=1}^J{1\over P_i^2}\Big)\Phi_{12\cdots J},\n
\Phi_{12\cdots J}&\equiv&Tr{\cal Y}_1{\cal Y}_2\cdots {\cal Y}_J ,
\eeqa
while the $J-1$ non-planar amplitudes are
\beqa
&&<j,p|\prod_{i=1}^J{1\over P_i^2}|j,p>_{np}\n
&\equiv &{n^{J+1}\over f^{4J}N}
\sum_{2\cdots J}\Big(\Phi_{23\cdots J1}^*
+(J-2)~cyclic~permutations \Big)
\Big(\prod_{i=1}^J{1\over P_i^2}\Big)\Phi_{12\cdots J} .
\eeqa

We can estimate non-analytic part of the planar amplitudes 
from the following recursive relation
\beqa
&&<j,p|\prod_{i=1}^J{1\over P_i^2}|j,p>_p\n
&=&{n\over f^4}\sum_{j_1,p_1,j',p'} 
\left\{
\begin{array}{ccc}
  j &  j_1 & j' \\
  l   &  l & l
 \end{array}\right\}^2
 \left\{
\begin{array}{ccc}
  p &  p_1 & p' \\
  l   &  l & l
 \end{array}\right\}^2\n
&&\times{(2j_1+1)(2p_1+1)(2j'+1)(2p'+1)\over j_1(j_1+1)+p_1(p_1+1)}
<j',p'|\prod_{i=2}^{J}{1\over P_i^2}|j',p'>_p .
\eeqa
Since
\beq
\sum_{j_1}(2j_1+1)\left\{
\begin{array}{ccc}
  j &  j_1 & j' \\
  l   &  l & l
 \end{array}\right\}^2
 ={1\over 2l+1} ,
 \eeq
and $|j_1-j'|\leq j$, we may evaluate it as 
\beqa
&&<j,p|\prod_{i=1}^J{1\over P_i^2}|j,p>_p\n
&\sim &
{n^2\over f^4N}\sum_{j'^2+p'^2>P^2} 
{(2j'+1)(2p'+1)\over j'(j'+1)+p'(p'+1)}
<j',p'|\prod_{i=2}^{J}{1\over P_i^2}|j',p'>_p\n
&&
+{n^2\over f^4N}\sum_{j'^2+p'^2<P^2} 
{(2j'+1)(2p'+1)\over P^2}
<j',p'|\prod_{i=2}^{J}{1\over P_i^2}|j',p'>_p .
\label{recurs} 
\eeqa
From this recursion relation, we identify the non-analytic part as
\beq
<j,p|\prod_{i=1}^J{1\over P_i^2}|j,p>_p
\sim 
(-1)^{J-2}{\lambda^{2J}\over (4\pi)^{2J}}
{1\over (J-1)!(J-2)!}(P^2)^{J-2}log({4l^2\over P^2}) ,
\label{plgr2p}
\eeq
corresponding to $1/x^{2J}$ behavior in real space.

As we have argued for the $J=2$ case, we should be able to reproduce these correlators
from those on the flat 4d space in the large $N$ limit. The planar amplitude can be estimated as
\beqa
I_J&\equiv&\lambda^{2J}\int \prod_1^{J} {d^4k_i\over (2\pi )^4}{1\over k_i^2}
(2\pi )^4\delta^4 (\sum_{i=1}^{J} k_i-k)\n
&=&\lambda^{2J}
\int {dx^4\over l^4} exp({1\over l}ik\cdot x)\Big({l^2\over 4\pi^2 (x^2+\delta^2 )}\Big)^J\n
&=&\lambda^{2J}{1\over \Gamma (J)}({1\over 4\pi^2})^Jl^{2(J-2)}
\int_0^{\infty} d \alpha \alpha^{J-1} \int dx^4 exp({1\over l}ik\cdot x-\alpha (x^2+\delta ^2 ))\n
&=&\lambda^{2J}\pi^2{1\over \Gamma (J)}({1\over 4\pi^2})^Jl^{2(J-2)}
\int_0^{\infty} d\alpha\alpha^{J-3}exp(-{k^2\over 4\alpha l^2}-\delta^2\alpha)\n
&=&2\lambda^{2J}\pi^2{1\over \Gamma (J)}({1\over 4\pi^2})^J
({l^2\over \delta^2})^{J-2}({\delta^2 k^2\over 4l^2})^{J-2\over 2}
K_{J-2}\Big(2\sqrt{{\delta^2 k^2\over 4l^2}}\Big) ,
\eeqa
where we have introduced the short distance cut-off $\delta$.
The non-analytic part can be identified as
\beq
I_J \sim {\lambda^{2J}\over (4\pi)^{2(J-1)}}{1\over \Gamma (J)\Gamma (J-1)}({-k^2})^{J-2}
log({4l^2\over k^2}) ,
\eeq
where we have put the short distance cut-off $\delta \sim 1$.
Alternatively we can obtain the identical result by performing the following partial integrations:
\beqa
&&{1\over 16\pi^2}(-k^2)^{J-2}log({l^2\over k^2})\n
&=&\int {d^4x\over  l^4}
({l^2\over 4\pi^2 x^2})^2({\partial^2})^{J-2}exp({1\over l}ik\cdot x)\n
&=&\Gamma (J)\Gamma (J-1)(4\pi)^{2(J-2)}
\int {d^4x\over l^4}exp({1\over l}ik\cdot x)({l^2\over 4\pi^2 x^2})^J .
\eeqa

The non-planar amplitudes can be estimated by using the
identical recursion relation with the planar amplitude case (\ref{recurs}).
However we need to use the different initial condition (\ref{J2np}) which contains
UV/IR mixing effect for the $J=2$ case.  
We can indeed argue that
the non-analytic behavior of this amplitude only comes from
such a part of the phase space.
In this way, we estimate each non-planar amplitude as
\beq
(-1)^{J-2}{\lambda^{2J}\over (4\pi)^{2J}}{1\over (J-1)!(J-2)!}(P^2)^{J-2}log({4l^2\over (P^2)^2}) .
\label{npgr2p}
\eeq
The non-planar contributions may also be estimated  in the flat limit:
\beq
I_{n,m}=\lambda^{2J}\int \prod_{i=1}^{J} {d^4k_i\over (2\pi )^4}{1\over k_i^2}
(2\pi )^4\delta^4
 (\sum_{i=1}^{J} k_i-k)exp({1\over l}ik \wedge \sum_{i=1}^m k_i) ,
\eeq
where $n+m=J$.
It is convenient to estimate it in real space as
\beqa
&&\lambda^{2J}\int {dx^4\over l^4} exp({1\over l}ik\cdot x)
\Big({l^2\over 4\pi^2 (x^2+\delta^2 )}\Big)^n\Big({l^2\over 4\pi^2 ((x-\tilde\delta)^2+\delta^2 )}\Big)^m\n
&=&\lambda^{2J}\pi^2{1\over \Gamma (n)\Gamma (m)}({1\over 4\pi^2})^{J}l^{2(J-2)}
\int_0^{\infty} d\alpha\alpha^{n-1}d\beta\beta^{m-1}
exp(-{k^2\over 4(\alpha+\beta )l^2}-\delta^2(\alpha+\beta) 
-\tilde\delta^2{\alpha\beta\over \alpha+\beta} )\n
&=&\lambda^{2J}\pi^2{1\over \Gamma (n)\Gamma (m)}({1\over 4\pi^2})^{J}l^{2(J-2)}
\int_0^1 d\alpha\alpha^{n-1}(1-\alpha )^{m-1}\int_0^{\infty} d\lambda\lambda^{J-3}\n
&&\times exp(-{k^2\over 4\lambda l^2}-\lambda\delta^2-\lambda\alpha(1-\alpha )\tilde\delta^2 ) ,
\label{np2pam}
\eeqa
where $|\tilde\delta |= |k| > \delta$.
The above expression is equal to
\beq
\lambda^{2J}\pi^2{1\over \Gamma (n)\Gamma (m)}({1\over 4\pi^2})^{J}
\int_0^1 d\alpha\alpha^{n-1}(1-\alpha )^{m-1}
2({l^2\over \bar\delta^2})^{J-2}({\bar\delta^2 k^2\over 4l^2})^{J-2\over 2}
K_{J-2}\Big(2\sqrt{{\bar\delta^2 k^2\over 4l^2}}\Big) ,
\label{modbes}
\eeq
where $\bar\delta^2=\delta^2+\alpha(1-\alpha )\tilde\delta^2$.
The singular part of the modified Bessel function behaves as
\beqa
&&K_{J-2}\Big(2\sqrt{{\bar\delta^2 k^2\over 4l^2}}\Big)\n
&=&{(-1)^{J-1}\over 2(J-2)!}\Big({\bar\delta^2 k^2\over 4l^2}\Big)^{J-2\over 2}
log({\bar\delta^2 k^2\over 4l^2})
+{(J-3)!\over 2}\Big({\bar\delta^2 k^2\over 4l^2}\Big)^{-{(J-2)\over 2}}
+\cdots ,
\eeqa
where $\cdots$ denotes higher order terms in $k^2$.
Since the second term in the last expression gives rise to
the short distance singularity as
\beq
\lambda^{2J}\pi^2({1\over 4\pi^2})^{J}
{1\over (J-1)(J-2)}
({l^2\over \delta ^2})^{J-2} +\cdots ,
\eeq
the  non-analytic behavior of the amplitude comes from the first term as
\beq
{\lambda^{2J}\over (4\pi)^{2(J-1)}}{1\over \Gamma (J)\Gamma (J-1)}
(-k^2)^{J-2}log({l^2\over (k^2)^2}) .
\label{np2pam1}
\eeq

We note that we could contemplate more generic extension of the chiral operators
in NC gauge theory on the flat 4d space such as
\beq
Tr exp({1\over l}i\alpha_1k\cdot A)Zexp({1\over l}i\alpha_2 k\cdot A)
Z\cdots exp({1\over l}i\alpha_J k\cdot A)Z ,
\label{gwilson}
\eeq
where $\sum \alpha_i =1$. 
This operator would correspond to $Tr {\cal Y}_{j_1j_1}z{\cal Y}_{j_2j_2}z\cdots 
{\cal Y}_{j_Jj_J}z$ on $S^2\times S^2$ where $\sum j_i=j$.
For a generic choice of $\alpha_i$, we can observe by power counting arguments that
the non-planar contributions to the 
two point correlators behave when $\delta \rightarrow 0$ as
\beq
\lambda^{2J}\int {dx^4\over l^4}exp({1\over l}ik\cdot x)\prod_{i=1}^J
{l^2\over 4\pi^2((x-\tilde\delta_i)^2+\delta^2)}
\sim \lambda^{2J}({l^2\over k^2})^{J-2} ,
\eeq
where $|\tilde\delta_i|=\alpha_i|k|$.
Since the planar contributions are identical to those in ordinary gauge theory,
such contributions completely alter the non-analytic behavior of the two point functions. 
We therefore restrict our considerations to $Tr  {\cal Y}z^J$ type operators in this paper.

By putting together planar and non-planar contributions:
(\ref{plgr2p}) and (\ref{npgr2p}), we obtain
\beqa
&&{n\over N}<Tr{\cal Y}_{j,p}z^JTr\bar{z}^J{\cal Y}_{j,p}^{\dagger}>\n
&\sim & (-P^2)^{J-2}log({l^2\over P^2})+(J-1)(-P^2)^{J-2}log({l^2\over (P^2)^2}) .
\eeqa
We can rewrite it as follows just like $J=2$ case in (\ref{2jtree})
\beq
J(-P^2)^{J-2}log({l\over P^2})+(J-1)(-P^2)^{J-2}log({l\over P^2})+(-P^2)^{J-2}log(l) .
\label{Jjtree}
\eeq
We find that the first term is identical to ordinary gauge theory with
UV cut-off of $\sqrt{l}$. The second term is specific to NC gauge theory representing
the long range interaction due to UV/IR mixing.
It is consistent to interpret it in terms of a Kaluza-Klein mode which couples to
this operator. The third term does not correspond to
long distance physics as it corresponds to the derivatives of $\delta$ function.

\section{Quantum corrections to the correlators}
\setcounter{equation}{0}

In this section, we investigate the quantum corrections to the two point correlators
of the Wilson lines at the one loop level. For this purpose, we need to consider
the renormalization of NC gauge theory on $S^2\times S^2$.
At the one loop level, we have found the divergence of the gauge field self-energy type
in the planar sector \cite{KTT}
\footnote{
We ignore mass and Chern-Simons type terms by focusing on large enough momentum scale.}. 
In order to remove it,
we may renormalize the gauge field as
\beqa
A_{\mu}&=&f(p_{\mu}+Z_{\Delta^2}a_{\mu}^{j,p}{\cal Y}_{j,p})\n
Z_{\Delta^2}&=&1-{1\over 8\pi^2}\lambda^2log\Big({N\over n\Delta^2}\Big) ,
\eeqa
where $\Delta$ is a renormalization scale.

This renormalization procedure does not remove the divergence
of the non-planar self-energy of the $SU(n)$ singlet field.
Such a divergence cannot be removed by a constant wave function renormalization of
the $SU(n)$ singlet field since it occurs only when $p^2 <l$.
This phenomenon is a manifestation of the $UV/IR$ mixing in NC gauge theory.
In other words, the gauge symmetry of NC gauge theory reduces to
ordinary gauge symmetry in the small momentum limit.
In such a limit, $U(n)$ gauge symmetry decouples as $SU(n)\times U(1)$.
It is therefore expected that the singlet field is renormalized differently
from the non-singlet fields in the small momentum regime.

We choose to let the divergent self-energy of the
singlet field in the small momentum regime as it is. 
We believe that this procedure does not spoil the renormalizability
of the theory since the singlet field decouples
from the non-singlet fields in the low momentum regime.
As a concrete example, we consider 
the two point functions with $J=1$ to which the singlet field contributes.
With generic momenta, they behave at the tree level as
\beq
{n\over N}<Tr{\cal Y}_{j,p}zTr\bar{z}{\cal Y}_{j,p}^{\dagger}>
={n\over f^4N}{1\over P^2} ,
\eeq
where $P^2=j(j+1)+p(p+1)$.
At the one loop level, we have found the following quantum corrections to this correlator
in our renormalization procedure when $P^2<<l$
\beqa
&&{n\over f^4N}\Big(Z_{P^2}^2
-\omega_{np} (P^2)\Big){1\over P^2}\n
&=&{1\over n}{\lambda^2\over (4\pi )^2}
\Big(1-{\lambda^2\over 4\pi^2}log(P^2)\Big){1\over P^2} ,
\label{j1cor}
\eeqa
where we have taken the renormalization scale $\Delta^2=P^2$
and $\omega_{np} (P^2)$ is the non-planar contribution to
gauge field self-energy.
The planar self-energy has been cancelled by the counter term with
this renormalization procedure.
We find that the correlator (\ref{j1cor}) requires no infinite
renormalization since it
is independent of $N$ with fixed 't Hooft coupling $\lambda^2$.
However we find a non-analytic finite correction.
Since we are dealing with non-local operators, there is always
a question as which operator corresponds to local operators.
We can indeed eliminate $\Big(1-\lambda^2log(P^2)/4\pi^2\Big)$
factor by redefining the $J=1$ operator.
\beq
Tr{\cal Y}z \rightarrow \Big(1-{\lambda^2\over 8\pi^2}log(P^2)\Big)Tr{\cal Y}z  .
\eeq
In fact such a freedom exists for the Wilson lines since the
operators with different momenta are independent of each other.
In perturbation theory,  we have found it possible to rescale
the operators with $J=1$ in such a way that
they can be interpreted as the Fourier 
transformation of a local operator with no anomalous dimensions.
We remark that there is no $J=1$ chiral operator in ordinary gauge theory
with $SU(n)$ gauge group to which AdS/CFT correspondence applies.
The supergravity description of this operator has been studied in
\cite{Tominosg}.

The most relevant chiral operator in ordinary gauge theory starts with $J=2$.
We move on to investigate the quantum corrections to the two point
correlators of the non-commutative analogue of the $J=2$ chiral operator.
The one loop self-energy corrections to the propagators are
\beqa
& & \epsfysize=15mm\parbox[c]{25mm}{\epsfbox{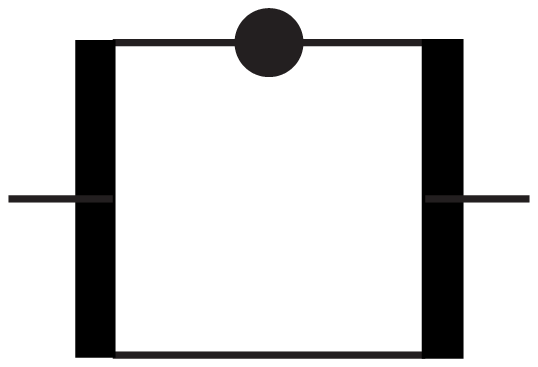}}
 +  \epsfysize=15mm\parbox[c]{25mm}{\epsfbox{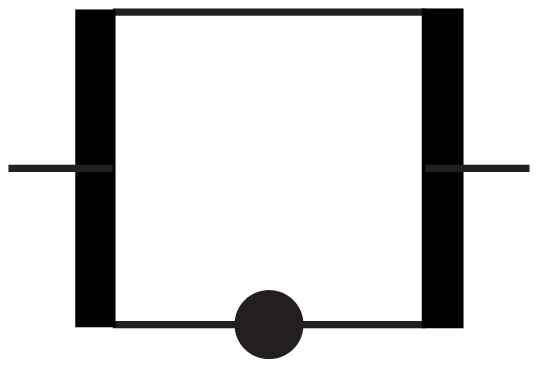}}\n
&+& \epsfysize=15mm\parbox[c]{25mm}{\epsfbox{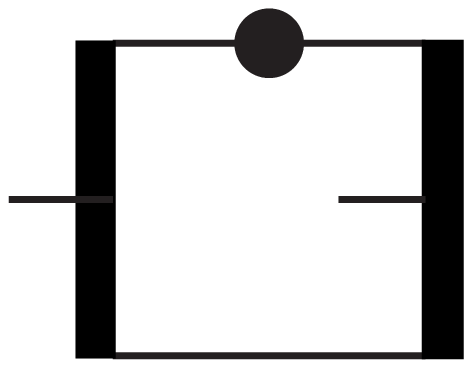}}
 +  \epsfysize=15mm\parbox[c]{25mm}{\epsfbox{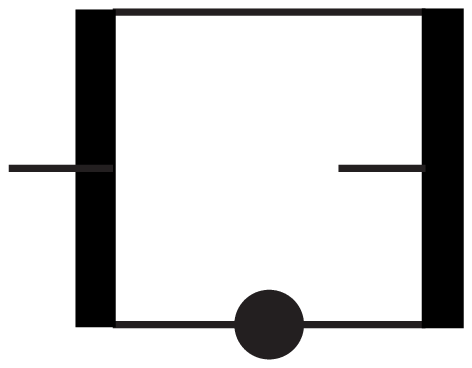}}\n
&=&-8 {n^4\over f^{12}N}\sum_{2345}\Psi_{123}^*\Psi_{\bar{2}45}^*
{1\over P_2^2P_3^2P_4^2P_5^2}\Psi_{\bar{2}45}\Psi_{123}\n
&&-8 {n^4\over f^{12}N}\sum_{2345}e^{i\phi_{123}}\Psi_{123}^*\Psi_{\bar{2}45}^*
{1\over P_2^2P_3^2P_4^2P_5^2}\Psi_{\bar{2}45}\Psi_{123} ,
\label{2Jqc1}
\eeqa
where the operators $P_i$ do not act on the
matrices labeled by $\bar{i}$ in our convention.
The other quantum corrections arise due to the following
interaction terms of the action.
\beq
Tr{1\over 2}[\bar{Z},Z]^2-[A_{\alpha},\bar{Z}][A^{\alpha},Z] .
\eeq
The quartic vertex gives rise to
\beqa
 &&\epsfysize=15mm{\parbox[c]{25mm}{\epsfbox{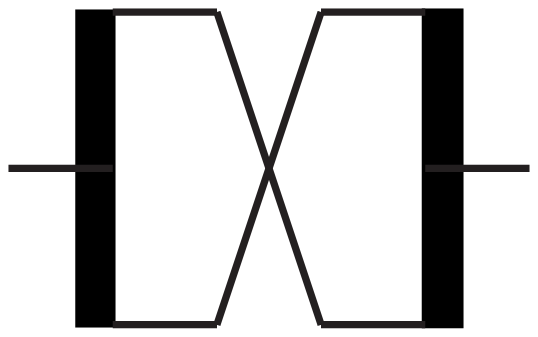}}}
+\epsfysize=15mm{\parbox[c]{25mm}{\epsfbox{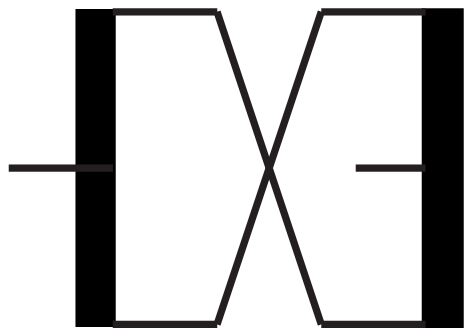}}}
+\epsfysize=15mm{\parbox[c]{25mm}{\epsfbox{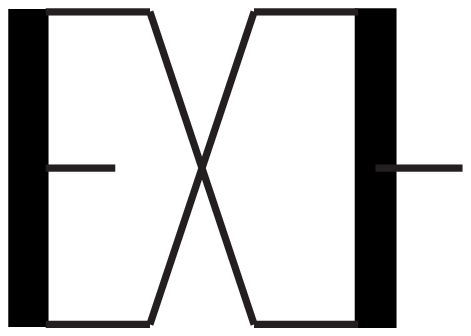}}}
+\epsfysize=15mm{\parbox[c]{25mm}{\epsfbox{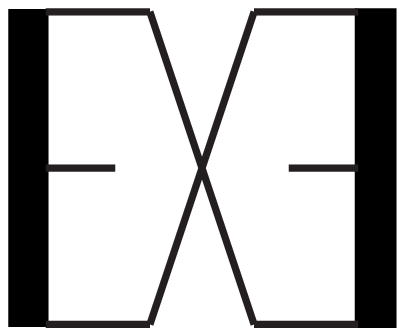}}}\n
&=&{n^4\over f^{12}N}
\sum_{2345}
{1\over P_2^2P_3^2P_4^2P_5^2}
\Psi_{45\bar{3}\bar{2}}\n
&&\times 
\Big(\Psi_{\bar{1}\bar{5}\bar{4}}\Psi_{123}
+\Psi_{\bar{1}\bar{5}\bar{4}}\Psi_{132}
+\Psi_{\bar{1}\bar{4}\bar{5}}\Psi_{123}+
\Psi_{\bar{1}\bar{4}\bar{5}}\Psi_{132}\Big) .
\eeqa
The cubic vertices give
\beqa
&&\epsfysize=15mm{\parbox[c]{25mm}{\epsfbox{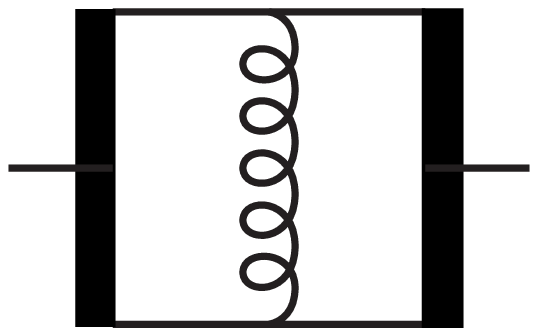}}}
+\epsfysize=15mm{\parbox[c]{25mm}{\epsfbox{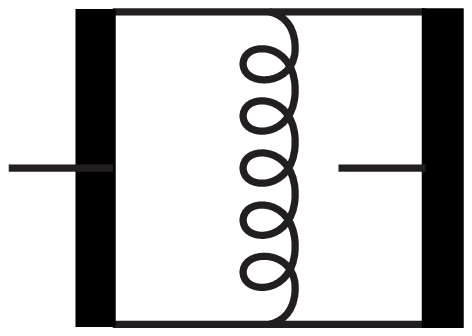}}}
+\epsfysize=15mm{\parbox[c]{25mm}{\epsfbox{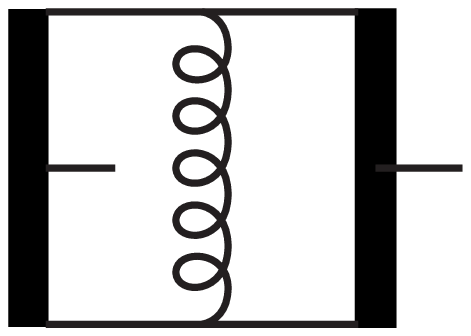}}}
+\epsfysize=15mm{\parbox[c]{25mm}{\epsfbox{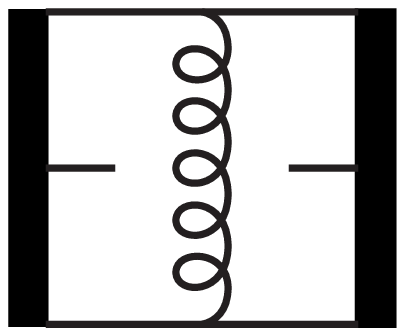}}}\n
&=&-{n^4\over f^{12}N}
\sum_{23456}
{(P_4+P_2)\cdot(P_3+P_5)\over P_2^2P_3^2P_4^2P_5^2P_6^2}
\Psi_{\bar{2}46}\Psi_{5\bar{3}\bar{6}}\n
&&\times 
\Big(\Psi_{\bar{1}\bar{5}\bar{4}}\Psi_{123}
+\Psi_{\bar{1}\bar{5}\bar{4}}\Psi_{132}
+\Psi_{\bar{1}\bar{4}\bar{5}}\Psi_{123}+
\Psi_{\bar{1}\bar{4}\bar{5}}\Psi_{132}\Big) .
\eeqa
By combing the above two, the leading contribution is found as
\beqa
&&{n^4\over f^{12}N}
\sum_{23456}
{ P_2+P_3^2+P_4^2+P_5^2\over P_2^2P_3^2P_4^2P_5^2P_6^2}
\Psi_{\bar{2}46}\Psi_{5\bar{3}\bar{6}}\n
&&\times 
\Big(\Psi_{\bar{1}\bar{5}\bar{4}}\Psi_{123}
+\Psi_{\bar{1}\bar{5}\bar{4}}\Psi_{132}
+\Psi_{\bar{1}\bar{4}\bar{5}}\Psi_{123}+
\Psi_{\bar{1}\bar{4}\bar{5}}\Psi_{132}\Big)\n
&=&4{n^4\over f^{12}N}
\sum_{2346}\Psi_{123}^*\Psi_{\bar{2}46}^*
{1\over P_2^2P_3^2P_4^2P_6^2}
\Psi_{\bar{2}46}\Psi_{123}\n
&&+4{n^4\over f^{12}N}
\sum_{2346}\Psi_{123}^*\Psi_{\bar{2}46}^*{1\over P_2^2P_3^2P_4^2P_6^2}
\Psi_{\bar{2}46}\Psi_{123}e^{i\phi_{123}}\n
&&+4{n^4\over f^{12}N}
\sum_{23456}{1\over P_2^2P_3^2P_4^2P_6^2}
\Psi_{\bar{2}46}\Psi_{5\bar{3}\bar{6}}\Psi_{\bar{1}\bar{5}\bar{4}}\Psi_{123}
e^{i\phi_{145}}\n
&&+4{n^4\over f^{12}N}
\sum_{23456}{1\over P_2^2P_3^2P_4^2P_6^2}
\Psi_{\bar{2}46}\Psi_{5\bar{3}\bar{6}}\Psi_{\bar{1}\bar{5}\bar{4}}\Psi_{123}
e^{i\phi_{123}}e^{i\phi_{145}} ,
\label{2j1l}
\eeqa
where $e^{i\phi_{123}}=(-1)^{j_1+j_2+j_3+p_1+p_2+p_3}$
and $e^{i\phi_{145}}=(-1)^{j_1+j_4+j_5+p_1+p_4+p_5}$.
Since the logarithmically divergent contribution 
in the first and second lines of the right-hand side in (\ref{2j1l}) cancels  only the half of (\ref{2Jqc1}), 
we have found that the two point function of the $J=2$ chiral operators
is  renormalized at the one loop level as
\beqa
&&-4{n^4\over f^{12}N}
\sum_{2346}\Psi_{123}^*\Psi_{\bar{2}46}^*
{1\over P_2^2P_3^2P_4^2P_6^2}
\Psi_{\bar{2}46}\Psi_{123}\n
&&-4{n^4\over f^{12}N}
\sum_{2346}\Psi_{123}^*\Psi_{\bar{2}46}^*{1\over P_2^2P_3^2P_4^2P_6^2}
\Psi_{\bar{2}46}\Psi_{123}e^{i\phi_{123}}\n
&&+4{n^4\over f^{12}N}
\sum_{23456}{1\over P_2^2P_3^2P_4^2P_6^2}
\Psi_{\bar{2}46}\Psi_{5\bar{3}\bar{6}}\Psi_{\bar{1}\bar{5}\bar{4}}\Psi_{123}
e^{i\phi_{145}}\n
&&+4{n^4\over f^{12}N}
\sum_{23456}{1\over P_2^2P_3^2P_4^2P_6^2}
\Psi_{\bar{2}46}\Psi_{5\bar{3}\bar{6}}\Psi_{\bar{1}\bar{5}\bar{4}}\Psi_{123}
e^{i\phi_{123}}e^{i\phi_{145}} .
\label{2j1lcr}
\eeqa

We are considering the case when the external momenta $P^2<<l$. 
In such a situation, we may adopt Edmonds'
approximation to compare $\Psi_{\bar{1}\bar{5}\bar{4}}$ and $\Psi_{\bar{1}\bar{4}\bar{5}}$.
After extracting the phase $e^{i\phi_{145}}$ from the $3j$ symbols, 
the former contains a factor
\beq
\left\{
\begin{array}{ccc}
  j &  j_4 & j_5 \\
  l   &  l & l
 \end{array}\right\}
\sim{(-1)^{j_5}\over \sqrt{(2j_5+1)(2l+1)}}d_{j_5-j_4,0}^{(j)}(\theta_5 ) ,
\eeq
while that of the latter is
\beq
\left\{
\begin{array}{ccc}
  j &  j_4 & j_5 \\
  l   &  l & l
 \end{array}\right\}(-1)^{j+j_4+j_5}
\sim{(-1)^{j_5}\over \sqrt{(2j_5+1)(2l+1)}}d_{j_5-j_4,0}^{(j)}(\pi-\theta_5 ) .
\eeq
The rotation matrices are given by the associated Legendre functions
\beq
d_{m,0}^{(j)}(\theta )
=\Big[{(j-m)!\over (j+m)!}\Big]^{1\over 2}P_j^m(cos(\theta )) .
\eeq
We would like to find out the condition when these two expressions
agree. 
For this purpose, we make use of the asymptotic behavior
of the associated Legendre functions for large $j$:
\beq
P_j^m(cos(\theta ))
\sim (-j)^m\sqrt{{2\over j\pi sin\theta }}
cos\Big( (j+{1\over 2}\Big)\theta +{m\pi\over 2}-{\pi\over 4}\Big) .
\eeq
From this expression, we can infer that the distance of the neighboring nodes
is $\pi/j$.
We argue that the planar and the non-planar amplitudes are coherent when
the difference of the arguments of the associated Legendre functions
($\theta_5$ and $\pi-\theta_5$) is smaller than it.
Such a requirement leads to the condition 
$j_4,j_5<\pi l/j$ and $p_4,p_5<\pi l/p$ as well.
This argument agrees with the estimate based on the effect of
non-commutative phase $exp(ik\wedge p/l)$ in the flat limit.

From these arguments, we can estimate (\ref{2j1lcr}) as
\beqa
&&-4{n^4\over f^{12}N}
\sum_{234'6} \Psi_{123}^*\Psi_{\bar{2}46}^*
{1\over P_2^2P_3^2P_4^2P_6^2}
\Psi_{\bar{2}46}\Psi_{123}\n
&&-4{n^4\over f^{12}N}
\sum_{234'6} \Psi_{123}^*\Psi_{\bar{2}46}^*{1\over P_2^2P_3^2P_4^2P_6^2}
\Psi_{\bar{2}46}\Psi_{123}e^{i\phi_{123}} ,
\eeqa
where $\sum_{234'6}$ implies that the summation is constrained 
with respect to $j_4,p_4$ as $l/j<j_4<l,l/p<p_4<l$.
With such a restriction, we can evaluate the following expression as
\beq
{4n\over f^{4}}
\sum_{4'6} \Psi_{\bar{2}46}^*
{1\over P_4^2P_6^2}
\Psi_{\bar{2}46}
\sim {\lambda^2\over 4\pi^2}log(P^2) .
\eeq
From these considerations, we conclude that the one loop correction
to the two point function of the $J=2$ chiral operators is
\beq
-{\lambda^2\over 4\pi^2}log(P^2) \times tree~result~(\ref{2jtree}) .
\eeq
Since it is proportional to a finite factor $log(P^2)$, 
this operator requires no infinite renormalization at the one loop level.
Nevertheless we find non-analytic finite corrections.
We may eliminate $\Big(1-\lambda^2log(P^2)/4\pi^2\Big)$
factor by redefining the $J=2$ operator just like $J=1$ case:
\beq
Tr{\cal Y}z^2 \rightarrow \Big(1-{\lambda^2\over 8\pi^2}log(P^2)\Big)Tr{\cal Y}z^2 .
\eeq
In perturbation theory,  we have found it possible again to rescale
the operators with $J=2$ in such a way that
they can be interpreted as the Fourier 
transformation of a local operator with no anomalous dimensions.

We can further argue that the leading one loop corrections to the correlators with $J>2$ are
just analogous to the $J=2$ case.
The one loop corrections in the planar sector with respect to gauge indices are local in the 
string world sheet sense as
they involve two neighboring $Z$ fields.
The chiral operators can be decomposed into $J$ sections 
each of which is bounded by two $Z$ fields.
Since each $Z$ field is shared by two sections,
the quarter of the self-energy correction to each $Z$ field can be associated with
a section.
In the commutative limit, such self-energy corrections are cancelled by
the rest of the one loop corrections in each section.
Therefore we only need to consider the one loop corrections which are affected by 
the non-commutative phase.  
Furthermore it is sufficient to consider the renormalization of
each operator.

We can thus focus on a section of an operator
which contains the spherical harmonics ${\cal Y}$. 
As the $J=2$ case, the one loop corrections
can be divided into the self-energy corrections and the rest. 
Were it not for ${\cal Y}$, they cancel each other.
With the presence of ${\cal Y}$, we need to interchange ${\cal Y}$ and
$Z$ field in this process.  After such an operation, the non-commutative phase could arise
which cuts-off the momentum integration.
Let us expand $Z$ field by matrices ${\cal Y}_1$.
We thus need to estimate the phase difference between
${\cal Y}{\cal Y}_1$ and ${\cal Y}_1{\cal Y}$.
For this purpose, we may compare
$Tr{\cal Y}{\cal Y}_1{\cal Y}_2$ and $Tr{\cal Y}_1{\cal Y}{\cal Y}_2$.
By the identical argument which was used to estimate the phase difference of 
$\Psi_{\bar{1}\bar{5}\bar{4}}$ and $\Psi_{\bar{1}\bar{4}\bar{5}}$,
we estimate that they are coherent when $j_1,j_2<\pi l/j$ and $p_1,p_2<\pi l/p$.
Since the upper momentum integration cut-off has been lowered in this way,
the renormalization factor differs by
the term $-{\lambda^2\over 8\pi^2}log(P^2)$
in comparison to the other sections. 
The two point functions of the generic chiral operators with $J \geq 2$ are thus 
renormalized as follows: 
\beq
-{\lambda^2\over 4\pi^2}log(P^2) \times tree~result~(\ref{Jjtree}) .
\label{qcr2j}
\eeq
We may eliminate the non-analytic finite factor $\Big(1-\lambda^2log(P^2)/4\pi^2\Big)$
by redefining the  $J\geq 2$ operators in general as:
\beq
Tr{\cal Y}z^J \rightarrow \Big(1-{\lambda^2\over 8\pi^2}log(P^2)\Big)Tr{\cal Y}z^J .
\eeq
We conclude that  we have found it possible to rescale
the operators with $J\geq 2$ in such a way that
they can be interpreted as the Fourier 
transformation of the local operators with no anomalous dimensions.

Although we can rescale the operators by momentum dependent factors to
fit any two point functions as we wish, such a procedure alters
multi-point functions. We argue that our procedure is legitimate since
it removes one loop quantum corrections of all correlators
since the renormalization effect is associated with individual
operators.
It is certainly necessary to make contact 
with supergravity since supergravity predicts vanishing anomalous dimensions
in the small momentum regime.

We can also study the following operators on $S^2\times S^2$.
\beq
y_{jm}^{\alpha_1,\alpha_2,\cdots,\alpha_j}
y_{pq}^{\beta_1,\beta_2,\cdots,\beta_p}
TrA_{\alpha_1}A_{\alpha_2}\cdots A_{\alpha_j}
A_{\beta_1}A_{\beta_2}\cdots A_{\beta_p}
\Phi_{i_1}\Phi_{i_2}\cdots ,
\eeq
where $\Phi_1=(A_6+iA_7)/\sqrt{2}$ and $\Phi_2=Z=(A_8+iA_9)/\sqrt{2}$.
We expand $A_{\mu}=f(p+a)_{\mu}$ as before and the leading terms are
\beq
Tr{\cal Y}_{jp}
\phi_{i_1}\phi_{i_2}\cdots ,
\eeq
where we have rescaled the operators to remove extra factors of $f$.
In ordinary gauge theory, these operators can be identified with the states of the spin chain
by associating $\phi_1$ with up spins and $\phi_2$ with down spins.

Since the one loop planar corrections with respect to gauge indices are local in the sense
that they involve adjacent $\phi$ fields, it is sufficient
to focus on a section of the operator.
We first consider the renormalization of
a section which contains no ${\cal Y}$.
Apart from the self-energy corrections of $\phi$ fields,  we need to compute
the quantum corrections due to the following
interaction terms of the action
\beqa
&&Tr\Big({1\over 2}[{\Phi_1}^{\dagger},\Phi_1]^2
+{1\over 2}[{\Phi_2}^{\dagger},\Phi_2]^2
-[A_{\alpha},{\Phi_i}^{\dagger}][A^{\alpha},\Phi_i]\n
&&-[{\Phi}_1^{\dagger},{\Phi}_2^{\dagger}][\Phi_1,\Phi_2]
-[{\Phi}_1^{\dagger},\Phi_2][\Phi_1,{\Phi}_2^{\dagger}]\Big) .
\eeqa
The quartic vertices give rise to
\beqa
&&-{n^3\over f^{12} N}
\sum_{45}
{1\over P_4^2P_5^2}\Psi_{145}
\Psi_{\bar{5}\bar{4}23}
(1-T)\n
&&+{n^3\over f^{12} N}
\sum_{45}
{1\over P_4^2P_5^2}\Psi_{145}
\Psi_{\bar{5}\bar{4}23}T ,
\eeqa
where $T$ is the exchange operator.
The cubic vertices give
\beq
-{n^3\over f^{12} N}
\sum_{456}
{(P_4+P_2)\cdot(P_3+P_5)\over P_4^2P_5^2P_6^2}\Psi_{145}
\Psi_{{2}6\bar{4}}\Psi_{{3}\bar{5}\bar{6}} .
\eeq
After including the quarter of the self-energy corrections
of $\phi$ fields which are associated with this section, 
we find the following quantum correction:
\beqa
&&-{2n^3\over f^{12} N}
\sum_{45}
{1\over P_4^2P_5^2}\Psi_{145}
\Psi_{\bar{5}\bar{4}23}
(1-T)\n
&=&-{n^2\over 2f^{8} N}\omega(P^2)
\Psi_{123}(1-T) .
\eeqa
This is identical to the ordinary gauge theory result giving rise to the anomalous dimension
of the spin chain Hamiltonian type \cite{MZ}\cite{BKS}\cite{KMMZ}:
\beq
H=J+{\lambda^2\over 8\pi^2}\sum_i(1-T_i)
=J+{\lambda^2\over 16\pi^2}\sum_i(1-\sigma_i\cdot\sigma_{i+1}) .
\eeq

In NC gauge theory, we need to study the renormalization effect
of the section which contains ${\cal Y}$ as well.
Just like the non-commutative chiral operator case, it gives rise to an extra 
renormalization factor
\beq
-{\lambda^2\over 8\pi^2}log(P^2)
+{\lambda^2\over 8\pi^2}log(P^2)(1-T_1) ,
\eeq
where $T_1$  exchanges the two fields
adjacent to ${\cal Y}$.
We can remove this factor by redefining the Wilson line operator $W$
\beq
W \rightarrow \Big(1-{\lambda^2\over 8\pi^2}log(P^2)
+{\lambda^2\over 8\pi^2}log(P^2)(1-T_1)\Big)
W .
\eeq
In this case also we can 
rescale the operators in such a way that
they can be interpreted as the Fourier 
transformation of the local operators 
whose anomalous dimensions are given by the spin chain Hamiltonian
of ordinary gauge theory.

So far we have neglected the gauge fields $a_{\mu}$ in the expansion
of $A_{\mu}$ around the classical solution $p_{\mu}$ as $A=f(a+p)_{\mu}$.
Before concluding this section, 
we investigate the effect of the gauge fields to the Wilson line correlators.
The leading correction to (\ref{yz}) is
\beqa
&&y_{j,j}y_{p,p}Tr\Big(\sum_k(p_{+})^{k}a_+(p_{+})^{j-k-1}
(\tilde{p}_{+})^{p}
+\sum_k(p_{+})^{j}(\tilde{p}_{+})^{k}\tilde{a}_+(\tilde{p}_{+})^{p-k-1}\Big)z^J\n
&\sim&{1\over l}Tr\Big(\sum_k{\cal Y}_{k,0}a_+{\cal Y}_{j-k-1,p}+
\sum_k{\cal Y}_{j,k}a_+{\cal Y}_{0,p-k-1}\Big)z^J .
\eeqa
From this structure, we can estimate the magnitude of
the gauge field effects to the two point functions as
\beq
\lambda^2{(P^2)^2\over l^2} .
\eeq
We can conclude that 
these corrections can be neglected as long as
the above quantity is small.
It implies that the gauge field effects can be neglected
when the operator probes the distance scale larger than $R$ where $R^4=\lambda^2l^2$.
We remark that $R$ coincides with the
radius of the background in dual supergravity
which will be studied in the next section.

\section{Supergravity description}
\setcounter{equation}{0}

It is an attractive idea that non-commutative gauge theories may also possess dual supergravity
descriptions \cite{hashimoto}\cite{ads+f}\cite{DhK}\cite{strsc}.
We have indeed demonstrated that we can perturbatively 
identify the non-commutative extensions of the BPS operators with no anomalous
dimensions in this paper. 
These observables with finite momenta probe the low energy limit of NC
gauge theory since the non-commutative scale is $O(N^{1\over 4})$.
In NC gauge theory, there are both IR and UV/IR contributions to the two point
correlators as we have seen in the preceding sections.
Since the IR contributions are identical to those in ordinary gauge theory, 
the supergravity background need not change from $AdS_5$ were it not for
UV/IR contributions.  We will argue in what follows that the way it deviates from  $AdS_5$
is consistent to accommodate UV/IR contributions.

We recall 
the Euclidean IIB supergravity action:
\beqa
S_{IIB}&=&S_{NS}+S_{R}+S_{CS},\n
S_{NS}&=&-{1\over 2}\int d^{10}x\sqrt{g}
e^{-2\phi}(R+4\partial_{\mu}\phi\partial^{\mu}\phi
-{1\over 2}{H_3}^2),\n
S_{R}&=&{1\over 4}\int d^{10}x\sqrt{g}
({F_1}^2+{\tilde{F}_3 }^2+{1\over 2}{\tilde{F}_5 }^2),\n
S_{CS}&=&{1\over 4}\int C_4\wedge H_3\wedge F_3 ,
\label{sugact}
\eeqa
where
\beqa
\tilde{F}_3&=&F_3-C_0\wedge H_3,\n
\tilde{F}_5&=&F_5-{1\over 2}C_2\wedge H_3+{1\over 2}B_2\wedge F_3 .
\eeqa

Supergravity solution which is dual to $U(n)$ $NCYM_4$ is
\beqa
e^{\phi}&=&
({ ng^2\over r^4})
{1\over (1+{ng  \over r^4})} ,\n
{1\over \alpha'}ds^2&=&
({ng\over r^4})^{1\over 2}(
{d\vec{x}^2\over
1+{ng  \over r^4}}+
dr^2 +r^2d\Omega_5^2 ) ,\n
B_2&= &{1\over (1+{ng \over r^4})}dx\wedge dy +
{1\over (1+{ng \over r^4})}dz\wedge d\tau ,\n
C_2&=&i{1\over g} B_2 ,\n
C_0&=& -i{r^4\over ng^2} ,\n
F_{0123r}&=&
-4i{1\over (1+{ng \over r^4})^2}{n\over r^5}
\label{adsrs} .
\eeqa
Here we have put NS $B$ field strength $b=1$ which implies that
the noncommutativity scale $l_{NC}$ is $O(1)$.
Since $S^2\times S^2$ approaches the flat 4d space in the large $N$ limit,
we believe that this solution is relevant to our problem.

Since $ng$ which corresponds to the 't Hooft coupling of $NCYM_4$
sets the radius of `$AdS_5$' and $S_5$
as $R^4/l_{NC}^4=ng$, supergravity description is expected to be valid
in the strong coupling limit.
It is because the mass scale for the Kaluza-Klein modes can be estimated to be of order
$1/R$ in comparison to that of the oscillator modes.
In NC gauge theory, we have observed in the preceding section
that the gauge field effects to the Wilson line correlators
are small if they probe the distance scale larger than  $R$ where $R^4=\lambda^2 l^2$.
Hence the identification of $ng$ and $\lambda^2$ is indeed 
consistent since the non-commutative scale is $l_{NC}\sim \sqrt{l}$ in NC gauge theory.
The supergravity regime corresponds to local field theory regime
in NC gauge theory.
We may further assume that $n$ is large
in order to keep the dilaton expectation value to be small.

It is useful to introduce the coordinate system where
the five dimensional subspace $(\vec{x},\rho)$ is conformally flat
\beq
{1\over \alpha'}ds^2=A(\rho )(d \vec{x}^2 +d
\rho^2)+R^2d\Omega_5^2.
\label{confmet}
\eeq
Since
\beq
\rho =\int_R^r dr\sqrt{1+{R^4\over r^4}},
\eeq
we find that
\beq
A(\rho )\sim R^2/\rho ^2 ,~~\rho \rightarrow \pm \infty .
\eeq
$A(\rho )$ has the unique maximum at $\rho=0$  ($r=R$).

For simplicity, we consider a
massless field $\varphi$ such as dilaton or graviton.
It is reasonable to assume that they provide us generic information
for the entire supergravity multiplets.  
Such a field $\varphi$ obeys the following equation of motion
\beq
{1\over 2}
\nabla^{\mu}\nabla_{\mu} \varphi
-\nabla^{\mu}\phi\nabla_{\mu}\varphi
=0.
\label{freeeq}
\eeq
Eq.(\ref{freeeq}) can be rewritten as:
\beqa
{H}\varphi&=&0,\n
{H}&=&-{A\over \sqrt{g}}\partial_{\mu}\sqrt{g}g^{\mu\nu}\partial_{\nu}
+2 Ag^{\mu\nu}\partial_{\mu}\phi\partial_{\nu}.
\label{Shrod1}
\eeqa
The Hamiltonian is
\beq
H=-(\vec{\nabla}^2+{\partial ^2 \over \partial
\rho^2} +({3\over 2}{A'\over A}-2\phi '){\partial\over \partial \rho}+
{A\over R^2}\hat{L}^2) ,
\label{Hamilton1}
\eeq
where $A'={\partial A/\partial \rho}$ and $\phi '={\partial \phi/\partial
\rho}$. The symbols $\vec{\nabla}^2$ and $\hat{L}^2$ denote
the Laplacians on the flat 4d space and $S^5$ respectively.

We concentrate on the $S$ wave of $S^5$ in what follows.
The eigenfunction of $H$ is found as $exp(ik\cdot x)\varphi$
with the eigenvalue $k^2+E$. 
The eigenvalue $E$ and its eigenfunction can be determined
by solving the following quantum mechanics problem.
\beq
-({\partial ^2 \over \partial
\rho^2} +({3\over 2}{A'\over A}-2\phi '){\partial\over \partial \rho}) \varphi
=E\varphi .
\label{eqnphi}
\eeq
We first investigate the behavior of the solutions in the asymptotic regions
$\rho \sim \pm \infty$. When $\rho \sim -\infty$,
the Hamiltonian is well approximated as
\beq
(-{\partial ^2 \over \partial \rho^2}
+{3\over \rho}{\partial\over\partial \rho}){\varphi}
=E {\varphi} .
\label{extsol}
\eeq
The solutions are given by the Bessel functions
${\rho}^2\sqrt{\omega}J_2(\omega\rho)$ and ${\rho}^2\sqrt{\omega}Y_2(\omega\rho)$
where $\omega^2=E$.
They behave as ${\varphi}\sim |\rho |^{3\over 2}e^{\pm i\omega\rho}$ for large
$|\rho |$.
In the asymptotic region $\rho \sim \infty$,
the Hamiltonian becomes
\beq
(-{\partial ^2 \over \partial \rho^2}
-{5\over \rho}{\partial  \over \partial \rho}) {\varphi}
=E {\varphi} .
\label{conmod}
\eeq
The solutions are again given by the Bessel functions
${\rho}^{-2}\sqrt{\omega}J_2(\omega\rho)$ and ${\rho}^{-2}\sqrt{\omega}Y_2(\omega\rho)$.
Such wave functions behave as
$\varphi \sim \rho^{-{5\over 2}}e^{\pm i\omega\rho}$ for $\rho \sim \infty$.

We may construct
the propagator as follows
\beq
G(x,y)=\sum_j <x|j>{1\over E_j}<j|y> ,
\label{defpro}
\eeq
where $|j>$ is the eigenstate of ${H}$ with the
eigenvalue $E_j$.
With $<\rho|j>={\rho}^2\sqrt{\omega}J_2(\omega\rho)$, 
the bulk propagator for negative $\rho$ (or small $r$)  which appears in ordinary $AdS$/CFT
correspondence is obtained:
\beqa
G(x,y)_B&=&{1\over R^8}
\int {d^4k \over (2\pi )^4}exp(i\vec{k}\cdot(\vec{x}-\vec{y}))
\int_0^{\infty} d\omega~\omega {1\over \vec{k}^2+\omega^2}\n
&&\times {\rho}^2J_2(\omega\rho ){\rho '}^2J_2(\omega\rho ') .
\label{adspro}
\eeqa
In fact it satisfies the desired equation in $AdS_5$
\beq
-{1\over \sqrt{g}}\partial_{\mu}\sqrt{g}g^{\mu\nu}
\partial_{\nu}G(x,y)_B={1\over \sqrt{g}}\delta (x-y) ,
\eeq
since the Bessel functions satisfy the following completeness condition:
\beq
\delta (\rho -\rho ')
=\int_0^{\infty} d\omega~\omega\sqrt{\rho}J_2(\omega\rho )\sqrt{\rho '}J_2(\omega\rho ') .
\eeq
We may reexpress the propagator as follows
\beqa
G(x,y)_B&=&{1\over 2R^8}
\int {d^4k \over (2\pi )^4}exp(i\vec{k}\cdot(\vec{x}-\vec{y}))
\int_{-\infty}^{\infty} d\omega~\omega {1\over \vec{k}^2+\omega^2}\n
&&\times{\rho_{<}^2}J_2(\omega\rho_{<} ){\rho_{>}^2}H^{(1)}_2(\omega\rho_{>} ) ,
\eeqa
where $\rho_{<}(\rho_{>})$ denotes the smaller (larger) quantity between $|\rho|$
and $|\rho '|$.

In this form, we can now pick the residue of the simple pole at $\omega=ik$
to estimate its long distance behavior
\beqa
G(x,y)_B&=&{1\over R^8}
\int {d^4k \over (2\pi )^4}
exp(i\vec{k}\cdot(\vec{x}-\vec{y})){\rho_{<}^2}I_2(k\rho_{<})
{\rho_{>}^2}K_2(k\rho_{>})\n
&\sim& {3\over 2\pi^2R^8}{\rho^4\rho '^4\over
(\vec{x}-\vec{y})^8} .
\label{lrintsg}
\eeqa
Alternatively it can be directly estimated 
from (\ref{adspro}) as
\beqa
&&\int d\omega~\omega {1\over \vec{k}^2+\omega^2}
 {\rho}^2J_2(\omega\rho ){\rho '}^2J_2(\omega\rho ')\n
 &\sim& {\rho^4\rho '^4\over 64}
 \int d\omega~\omega^5 {1\over \vec{k}^2+\omega^2}\n
 &\sim& -{\rho^4\rho '^4\over 128}(k^2)^2logk^2 ,
\eeqa
where we have retained the non-analytic part in $k^2$ which reproduces
the long range interaction (\ref{lrintsg}) after the Fourier transformation.

The bulk propagator for positive  $\rho$ (or large $r$)  
is obtained in an analogous way:
\beqa
G(x,y)_B&=&
\int {d^4k \over (2\pi )^4}exp(i\vec{k}\cdot(\vec{x}-\vec{y}))
\int_0^{\infty} d\omega~\omega {1\over \vec{k}^2+\omega^2}\n
&&\times {\rho}^{-2}J_2(\omega\rho ){\rho '}^{-2}J_2(\omega\rho ')\n
&=&
\int {d^4k \over (2\pi )^4}
exp(i\vec{k}\cdot(\vec{x}-\vec{y})){1\over \rho_{<}^2}I_2(k\rho_{<})
{1\over \rho_{>}^2}K_2(k\rho_{>})\n
&\sim &{3\over 2\pi^2}{1\over (\vec{x}-\vec{y})^8} ,
\label{ncpro}
\eeqa
where $\rho_{<}(\rho_{>})$ denotes the smaller (larger) quantity between $\rho$
and $\rho '$.
It is identical to the propagator in the flat 10d spacetime.

In $AdS$/CFT correspondence in ordinary gauge theory,
the following relation plays an important role.
\beqa
&& \int {d^4k \over (2\pi )^4}
exp(i\vec{k}\cdot(\vec{x}-\vec{y})){\rho_{<}^2}I_2(k\rho_{<})
{\rho_{>}^2}K_2(k\rho_{>})\n
&\sim &
{1\over 4}{\rho_{<}^4}
\delta^4(\vec{x}-\vec{y}) ,
\eeqa
when $\rho_{>},\rho_{<}\rightarrow 0$. 
Hence we can construct the classical solution $\phi(\rho,\vec{x})$ 
which approaches $\phi(\vec{x})$ as 
$\rho$ approaches the boundary $\rho_0\sim 0$:
\beqa
\phi(\rho,\vec{x})&=&\int d^4y{4\over \rho_{0}^4}\int {d^4k \over (2\pi )^4}
exp(i\vec{k}\cdot(\vec{x}-\vec{y})){\rho^2}K_2(k\rho)
{\rho_{0}^2}I_2(k\rho_{0})\phi (\vec{y})\n
&=&\int {d^4k \over (2\pi )^4}exp(i\vec{k}\cdot\vec{x})
{4\over \rho_{0}^4}{\rho^2}K_2(k\rho)
{\rho_{0}^2}I_2(k\rho_{0})\phi (\vec{k}) .
\eeqa
This relation is used to estimate the boundary
contribution of the supergravity action.
It is a crucial ingredient to reproduce the correlators of local 4d field theory on the boundary.
However we are effectively constrained such that $|\rho|>R$ in NC gauge theory.
It is because our approximations through Bessel functions are no longer valid
when $|\rho|\rightarrow R$. We will verify the existence of the effective 
cut-off $R$ by an exact analysis through Mathieu functions subsequently.
Nevertheless there must be a supergravity description of NC gauge theory which
reproduces the correlators in the low momentum regime. 
We will argue in what follows that such a prescription is to
locate the Wilson lines at the maximum of the string metric $r\sim R$
\cite{DhK}.

After the change of the variables as
\beq
r=Re^{-z}, \varphi = {1\over r^2}\psi (z) ,
\eeq
$\psi $ obeys
\beq
[-{\partial ^2\over \partial z^2}-2(\omega R)^2cosh 2z]\psi (z) =-4\psi (z) .
\label{mateqn}
\eeq
(\ref{mateqn}) is identical to the wave equation of scalar fields in the background of
the D3-brane metric.
Therefore the exact propagator of this problem may be constructed through Mathieu functions.
It is interesting to note that we obtain the identical equation of the
motion in the large $b$ scaling limit with the D3-brane background without $b$ field.
In this coincidence the non-commutative scale in the former plays the role of the string scale
in the latter.  It is consistent with the proposal that NC gauge theory
is a string theory whose string scale is non-commutative scale \cite{strsc}.

The long distance behavior of the propagator is determined by the wave functions
with small $\omega$ since we will eventually pick a pole at $\omega =ik$.
In the small $\omega$ limit, our quantum system is separated into
two sectors whose wave functions are localized around $z\sim -\infty$ and $z \sim \infty$ 
due to the large potential energy barrier in the above expression.
Therefore our quantum system decouples into two sectors in the low energy limit.
It is therefore makes a sense to separate the system into two regimes:
$r<R$ regime and $r>R$ regime. 
It is consistent to locate the Wilson lines at the maximum 
of the metric $r=R$ with such a separation. 

In what follows, we evaluate the supergravity action 
for a classical solution with the fixed value $\phi(\vec{k})$ at $r=R$.
In Appendix (\ref{clssol}) , we have constructed a classical solution which approaches
$\phi(\vec{k})$ at $r=R$ in the small momentum regime.
The Wilson lines separate the $r<R$ region from the $r>R$ region.
With this classical solution, we can evaluate the 
contribution from $r<R$ region by the supergravity action (\ref{sugact})
just like ordinary gauge theory:
\beqa
&& {\pi^3R^5\over 2g^2}\int {d^4k \over (2\pi )^4}
 \phi (r,R,\vec{k}){\partial \over \partial  r}\phi (r,R,\vec{k})|_{r=R}\n
&=&{\pi^3R^8\over 2g^2}\int {d^4k \over (2\pi )^4}
\phi (\vec{k})\Big(-{1\over 8}k^4log(k^2R^2)\Big)\phi (-\vec{k})\n
&=&{n^2\over 16 \pi^2}\int {d^4k \over (2\pi )^4}
\phi (\vec{k})\Big(-{1\over 8}k^4log(k^2R^2)\Big)\phi (-\vec{k}) .
\eeqa
We can also evaluate the contribution from $r>R$ region:
\beqa
&& -{\pi^3R^5\over 2g^2}\int {d^4k \over (2\pi )^4}
 \phi (R,r',\vec{k}){\partial \over \partial  r'}\phi (R,r',\vec{k})|_{r'=R}\n
&=&{\pi^3R^8\over 2g^2}\int {d^4k \over (2\pi )^4}
\phi (\vec{k})\Big(-{1\over 8}k^4log(k^2R^2)\Big)\phi (-\vec{k})\n
&=&{n^2\over 16 \pi^2}\int {d^4k \over (2\pi )^4}
\phi (\vec{k})\Big(-{1\over 8}k^4log(k^2R^2)\Big)\phi (-\vec{k}) .
\label{bacnpl}
\eeqa
The sign difference between the above two expressions originates from the fact that
the former picks up the upper boundary
contribution while the latter picks up the lower boundary contribution
with respect to $r$ integration.
The both give the identical contributions to the two point function
in agreement with the propagator itself  at $r=R$ which is
evaluated in Appendix (\ref{exlepr}). 

We recall here that the two point functions of the Wilson lines
receive contributions from the planar and non-planar sectors. 
The non-analytic behavior arises not only from the small 
but also from large momentum contributions
due to UV/IR mixing in the non-planar sector.  The both contributions result in
the identical non-analytic behavior of the correlators
leading to the identical long range interaction. 
The planar contributions are identical to ordinary gauge theory while the non-planar contributions
are of the same magnitude.
In supergravity, we have also two decoupled sectors in the small momentum limit: 
$r<R$ and $r>R$ regimes.
We find that the supergravity action in these two different regimes
can account for the identical non-analytic behaviors.
It is reassuring that we can reproduce these essential features of the Wilson line correlators
in NC gauge theory from supergravity.
It a posteriori justifies our interpreting the contributions from $r<R$ and $r>R$ regimes in supergravity
as the IR and UV/IR mixing contributions in NC gauge theory respectively.
In fact the non-analytic part of the propagator near $r=R$ (\ref{exlepr}) is the 
sum of (\ref{lrintsg})  for $r<R$ and (\ref{ncpro}) for $r>R$ regimes.
Our prescription in supergravity is successful to describe this important feature
of two point functions of NC gauge theory.

\section{Conclusions and Discussions}
\setcounter{equation}{0}

In this paper, we have investigated the two point correlation functions
of the Wilson lines in NC gauge theory.
We have focused on ${\cal N}=4$ gauge theory on $S^2\times S^2$
which is realized by IIB matrix model. 
We have found finite quantum corrections to the non-commutative 
extension of BPS operators which carry finite momenta.
We have further given a perturbative prescription to obtain local
operators with no anomalous dimensions in the small momentum regime.
We have argued that our prescription is legitimate as it removes one loop quantum
corrections of all correlators since the renormalization effect is associated with 
individual operators.

It has been conjectured that these correlators are described by dual supergravity
in the strong 't Hooft coupling regime.
Our findings in this paper summarized above support such a conjecture.
We find extra contributions to the correlators due to the UV/IR mixing
effect in addition to the identical IR contributions with
ordinary gauge theory.
We can successfully reproduce these characteristic features of NC gauge theory
by locating the Wilson lines at the maximum of the string metric in dual supergravity
description.

In $AdS$/CFT correspondence, ordinary 4d gauge theory is located at the boundary
of $AdS_5$ where the metric diverges. Since an arbitrary small distance in field theory
corresponds to a finite physical distance in such a situation, it is consistent to
propose that a field theory holographically realizes supergravity and string theory.
However the metric does not diverge anywhere in the supergravity background
(\ref{adsrs})  which is relevant to NC gauge theory. 
This feature is consistent with the fact that
the both NC gauge theory and string theory have minimum length scale.
By locating the NC gauge theory at the maximum of the string metric,
we can probe the shortest length scale in NC gauge theory for a fixed
supergravity length scale .
In more generic spacetime such as flat or de Sitter spacetime, the metric does not
diverge either.
We thus believe that supergravity background dual to NC gauge theory
has much in common with physically realistic spacetime.

In the one loop effective action of NC gauge theory constructed from
IIB matrix model, the bilinear terms of the Wilson lines
appear due to the non-planar diagrams \cite{IKK}.  
The leading terms are of the same type with (\ref{bacnpl})
after replacing $n\phi$ by the Wilson lines.  
Since such terms arise due to graviton
exchange in the bulk, 
they constitute an evidence for
the existence of dynamical supergravity in NC gauge theory
with finite $n$.
It is thus conceivable that non-planar sectors may  be effectively described by
dynamical supergravity. Such a possibility is certainly the most attractive feature of matrix models.
We would like to extend our scope of research to such a problem
by investigating finite $n$ corrections.

We can further mention that $\varphi=1$ is a solution of 
(\ref{mateqn}) with $\omega=0$.
It has been pointed out that this zeromode could give rise to
$4d$ gravity a la Randall and Sundrum \cite{strsc}.
We have shown that it makes sense to divide supergravity action into two
sectors: $r<R$ and $r>R$ regimes.
Let us substitute $\varphi=\varphi (\vec{x})$ into the supergravity action
in the $r<R$ regime. We obtain the following contribution:
\beq
{1\over 8g^2}\int_0^{R} dr r^5d^4x \partial_i\varphi(\vec{x})\partial_i\varphi(\vec{x})
(1+{R^4\over r^4})
= {3R^6 \over 16g^2}\int d^4x \partial_i\varphi(\vec{x})\partial_i\varphi(\vec{x}) ,
\eeq
which gives rise to massless fields in 4 dimensions.
Suppose $\varphi(\vec{x})$ is coupled to energy momentum tensor $T$ 
at $r=R$ as
\beq
\int d^4x T(\vec{x})\varphi (\vec{x}) .
\eeq
We then obtain 4d gravity with Newton's law whose gravitational coupling
constant is $R^2/n^2$.
However 
we have not found the direct evidence for this phenomenon
in our perturbative analysis in this paper.
Presumably this very interesting possibility may be realized
through non-perturbative effects in NC gauge theory.

\begin{center} \begin{large}
Acknowledgments
\end{large} \end{center}
This work is supported in part by the Grant-in-Aid for Scientific
Research from the Ministry of Education, Science and Culture of Japan.
A part of this work was carried out while one of us (Y.K.) visited 
The Banff International Research Station.
He thanks the organizers of the workshop 
and especially G. Semenoff for his hospitality.

\section*{Appendix A}
\renewcommand{\theequation}{A.\arabic{equation}}
\setcounter{equation}{0}

In this Appendix, we construct a classical solution in the bulk which approaches
$\phi(\vec{k})$ at $r=R$
in the small momentum regime by Mathieu functions \cite{GH}
\footnote{We use the identical notations with that reference.}.
The two independent solutions of (\ref{mateqn}) can be chosen to be the Floquet solutions:
\beq
J(\nu,z),~J(-\nu,z) .
\eeq
Here the parameter $\nu$ is determined in terms of the combination $\omega R$.
It has a power series expansion given by
\beq
\nu=2-{i\sqrt{5}\over 3}({\omega R\over 2})^4+
{7i\over 108\sqrt{5}}({\omega R\over 2})^8+\cdots .
\eeq
For our purpose, it is more appropriate to consider
\beqa
H^{(1)}(\nu,z)&=&{2\over C}\Big(J(-\nu,z)-{1\over \eta}J(\nu,z)\Big)\n
H^{(2)}(\nu,z)&=&-{2\over C}\Big(J(-\nu,z)-{\eta}J(\nu,z)\Big) ,
\eeqa
where $\eta=exp(i\pi\nu )$ and $C=\eta-1/\eta$.
The Mathieu functions approach respective Bessel functions as 
$Re z\rightarrow \infty$:
\beq
Z^{(j)}(\nu,z)\rightarrow Z_{\nu}^{(j)}(\sqrt{q}e^z) ,
\label{asympt}
\eeq
where $Z$ denotes $J$ or $H$.

The exact propagator due to excited states has been constructed in \cite{DG}
\beqa
G(x,y)_B&=&
\int {d^4k \over (2\pi )^4}exp(i\vec{k}\cdot(\vec{x}-\vec{y}))\n
&&({1\over rr '})^2 {C\over 2A}
H^{(2)}(\nu, -z'-{i\pi\over 2})|_{\omega =k}
{\pi \over 2i}H^{(1)}(\nu,z+{i\pi\over 2})|_{\omega =k} .
\label{daspro}
\eeqa
In the small $r,r'$ regime  ($z>z'>0$),  we can evaluate it as
\beqa
G(x,y)_B&=&\int {d^4k \over (2\pi )^4}
exp(i\vec{k}\cdot(\vec{x}-\vec{y}))  \n
&&({1\over rr '})^2
\Big(-J(\nu, z'+{i\pi\over 2})
+{\eta C\over 2A\chi}H^{(1)}(\nu,z'+{i\pi\over 2})\Big)|_{\omega =k}\n
&&\times {\pi \over 2i}H^{(1)}(\nu,z+{i\pi\over 2})|_{\omega =k} .
\eeqa
In the large $r,r'$ regime ($-z'>-z>0$),  it can be evaluated as
\beqa
G(x,y)_B&=&\int {d^4k \over (2\pi )^4}
exp(i\vec{k}\cdot(\vec{x}-\vec{y}))\n
&&\times ({1\over rr '})^2
\Big(J(\nu, -z-{i\pi\over 2})
+{C\over 2A\chi \eta }H^{(2)}(\nu,-z-{i\pi\over 2})\Big)|_{\omega =k}\n
&&\times {\pi \over 2i}H^{(2)}(\nu,-z'-{i\pi\over 2})|_{\omega =k} .
\eeqa
From the asymptotic behaviors in (\ref{asympt}), we can see that
this propagator indeed approaches (\ref{lrintsg}) when $z,z'\rightarrow \infty$
and (\ref{ncpro}) when $z,z'\rightarrow -\infty$.

The long range behavior of the propagator can be estimated by its
small momentum behavior. 
In such a limit, the following quantities behave as
\beqa
&&\chi \sim (-{2\over 3}-i{\sqrt{5}\over 3})
\Big(1+i{2\sqrt{5}\over 3}({kR\over 2})^4log({kR\over 2})\Big)
={\chi}_0+O(k^4),\n
&&A\equiv\chi-{1\over \chi}=-2i{\sqrt{5}\over 3}\Big(
1-({\chi}_0+{1\over {\chi}_0})({kR\over 2})^4log({kR\over 2})\Big)
= A_0+O(k^4),\n
&&\eta\sim 1+{\pi\sqrt{5}\over 3}({kR\over 2})^4,
~C=\eta-{1\over \eta}\sim {2\pi\sqrt{5}\over 3}({kR\over 2})^4 .
\eeqa
The Mathieu functions can be expanded as
\beqa
&&J(\nu, z+{i\pi\over 2})|_{\omega =k}\n
&\sim& -{1\over 2}({kR\over 2})^2
({R^2 \over r^2}+{1\over {\chi}_0}{r^2\over R^2})
\Big(1-i{\sqrt{5}\over 3} ({kR\over 2})^4log ({kR\over 2}{R\over r})\Big),\n
&&J(-\nu, z+{i\pi\over 2})|_{\omega =k}\n
&\sim& -{1\over 2}({kR\over 2})^2
({R^2 \over r^2}+{\chi}_0{r^2\over R^2})
\Big(1+ i{\sqrt{5}\over 3}({kR\over 2})^4log ({kR\over 2}{R\over r})\Big),\n
&&H^{(1)}(\nu, z+{i\pi\over 2})|_{\omega =k}\n
&\sim&
-{A_0\over C}({kR\over 2})^2{r^2\over R^2}
\Big(1-{1\over 2} ({kR\over 2})^4
\Big(2{R^4\over r^4}+({\chi}_0+{1\over {\chi}_0})\Big)
 log ({kR\over 2}{R\over r})\Big),\n
 &&H^{(2)}(\nu, -z-{i\pi\over 2})|_{\omega =k}\n
& \sim&{A_0\over C}({kR\over 2})^2{R^2\over r^2}
\Big(1- {1\over 2}({kR\over 2})^4
\Big(2{r^4\over R^4}+({\chi}_0+{1\over {\chi}_0})\Big)
 log ({kR\over 2}{r\over R})\Big) .
\eeqa
In this way we find the integrand of the propagator in (\ref{daspro}) behaves as
\beqa
&&({1\over rr '})^2 {C\over 2A}
H^{(2)}(\nu, -z'-{i\pi\over 2})|_{\omega =k}
{\pi \over 2i}H^{(1)}(\nu,z+{i\pi\over 2})|_{\omega =k}\n
&&\sim {1\over 4}({1\over r'})^4
\Big(1-({kR\over 2})^4({r'^4\over R^4}+{R^4\over r^4})log ({kR\over 2})\Big) .
\label{exlepr}
\eeqa

We can also determine the small momentum expansion of the classical solutions as follows
\beqa
\phi (r,R,\vec{k})&\sim&
\Big(1-({kR\over 2})^4({R^4\over r^4}-1)log ({kR\over 2})\Big)
\phi(\vec{k}),~(r<R) ,\n
\phi (R,r',\vec{k})&\sim&{R^4\over {r'}^4}
\Big(1-({kR\over 2})^4({r'^4\over R^4}-1)log ({kR\over 2})\Big)
\phi(\vec{k}),~(R<r') ,
\label{clssol}
\eeqa
where the overall $k$ dependent normalization is fixed by the requirement that
they coincide $\phi(\vec{k})$ when $r,r'\rightarrow R$.

The propagator in (\ref{exlepr}) vanishes in the both asymptotic regions when $r\rightarrow 0$ or
$r'\rightarrow \infty$. When we locate the Wilson lines at $r=R$, such a requirement may be too restrictive.
For example, the following term can be added to it in the $r,r'>R$ regime
with an arbitrary coefficient since it solves the equation of motion:
\beq
({1\over rr '})^2 
{\pi \over 2i}H^{(2)}(\nu, -z'-{i\pi\over 2})|_{\omega =k}
{\pi \over 2i}H^{(2)}(\nu,-z-{i\pi\over 2})|_{\omega =k}
\sim 
{1\over 4}({R\over rr'})^4({2\over kR})^4 .
\eeq
Such a non-analytic behavior may explain the non-planar contributions to 
the two point correlators 
of more generic Wilson line operators we have mentioned in (\ref{gwilson}).

\end{document}